\documentclass[alpha-refs]{wiley-article}

\usepackage{placeins}
\usepackage{graphicx}
\usepackage[space]{grffile}
\usepackage{latexsym}
\usepackage{textcomp}
\usepackage{longtable}
\usepackage{tabulary}
\usepackage{booktabs,array,multirow}
\usepackage{amsfonts,amsmath,amssymb}
\usepackage{natbib}
\usepackage{url}
\usepackage{hyperref}
\hypersetup{colorlinks=false,pdfborder={0 0 0}}
\usepackage{etoolbox}
\makeatletter
\patchcmd\@combinedblfloats{\box\@outputbox}{\unvbox\@outputbox}{}{%
  \errmessage{\noexpand\@combinedblfloats could not be patched}%
}%
\makeatother
\newif\iflatexml\latexmlfalse

\AtBeginDocument{\DeclareGraphicsExtensions{.pdf,.PDF,.eps,.EPS,.png,.PNG,.tif,.TIF,.jpg,.JPG,.jpeg,.JPEG}}

\usepackage[utf8]{inputenc}
\usepackage[english]{babel}

\usepackage{siunitx}
\usepackage{algorithmic} %For computer algorithm code in LATEX
\usepackage{algorithm} %For algorithm box around the algorithmic
\usepackage{rotating} %For rotating tables (provides sidewaysfigure and sidewaystable)
\usepackage{lscape}

\newcommand{\iid}{\stackrel{\mathrm{iid}}{\sim}}

\newcommand{\amin}[1]{\underset{#1}{\operatorname{argmin~}}} %~ gives spacing
\newcommand{\reals}{\mathbb{R}}

\newtheorem{assumption}{Assumption}

\iflatexml
\else
\paperfield{Field of the paper}
\corraddress{Hyunseung Kang, Department of Statistics, University of Wisconsin-Madison, Madison, WI, 53706, U.S.A.}
\corremail{hyunseung@stat.wisc.edu}
\fundinginfo{Hyunseung Kang, Department of Statistics, Grant/Award Number: NSF DMS 1811414}
\fi

\papertype{Original Article}

\title{Inferring Treatment Effects After Testing Instrument Strength in Linear Models}

\author[1]{Nan Bi}
\author[2]{Hyunseung Kang}
\author[1]{Jonathan Taylor}

\affil[1]{Department of Statistics, Stanford University}
\affil[2]{Department of Statistics, University of Wisconsin-Madison}

\runningauthor{N. Bi, H. Kang, and J. Taylor}

\begin{document}

\maketitle
\selectlanguage{english}
\begin{abstract}
%A common practice in instrumental variables (IV) studies is to test instrument strength with an F-test and proceed to test the treatment effect if the F-statistic is above some threshold. However, many tests for the treatment effect do not account the strength test done a priori. In this paper, we show that this common practice leads to over-inflated Type I errors. We propose a method to correct it by leveraging recent advances in selective inference. We conclude by reanalyzing the effect of education on earning and show that not accounting for strength testing can dramatically alter the original conclusions. 
A common practice in IV studies is to check for instrument strength, i.e. its association to the treatment, with an F-test from regression. If the F-statistic is above some threshold, usually 10, the instrument is deemed to satisfy one of the three core IV assumptions and used to test for the treatment effect. However, in many cases, the inference on the treatment effect does not take into account the strength test done a priori. In this paper, we show that not accounting for this pretest can severely distort the distribution of the test statistic and propose a method to correct this distortion, producing valid inference. A key insight in our method is to frame the F-test as a randomized convex optimization problem and to leverage recent methods in selective inference. We prove that our method provides conditional and marginal Type I error control. We also extend our method to weak instrument settings. We conclude with a reanalysis of studies concerning the effect of education on earning where we show that not accounting for pre-testing can dramatically alter the original conclusion about education's effects.
%\textbf{Keywords} --- F-test, Instrument strength, Instrumental variables, Selective inference, Weak instruments
\end{abstract}%

\section{Introduction}
\subsection{Motivation}
Instrumental variables (IV) is a commonly used approach in economics, epidemiology, genetics, and health policy to estimate the effect of an exposure, treatment, or policy on an outcome of interest; see \citet{angrist_instrumental_2001}, \citet{hernan_instruments_2006}, and \citet{baiocchi_instrumental_2014} for overviews. IV methods require finding variables, known as instruments, that are (A1) associated with the treatment variable, (A2) unrelated to unmeasured confounders affecting both the outcome and the treatment, and (A3) have no direct effect on the outcome. %Assumptions (A2) and (A3) are sometimes combined into one assumption where the instruments are exogenous of the error terms in a structural model \citep{angrist_identification_1996,wooldridge_econometrics_2010}. 
The focus of the paper is on assumption (A1), specifically testing the treatment effect parameter $\beta^* \in \reals$, say $H_0: \beta^* = \beta_0$, with a test statistic $T$ \emph{after} testing for (A1), commonly referred to as testing for instrument strength. The goal can be roughly stated as follows
\begin{equation} \label{eq:1}
P_{H_0} (T \geq t \mid \text{tested IV strength}), \quad{} t \in \reals, 
\end{equation}
We assume throughout that (A2) and (A3) hold, even though these assumptions are important to assess in practice \citep{hernan_instruments_2006,conley2012plausibly,baiocchi_instrumental_2014,kang_instrumental_2016}.

The motivation for studying \eqref{eq:1} comes from a common practice in IV studies to test for instrument strength before testing the treatment effect. In particular, when instruments are weakly associated with the treatment and therefore poorly satisfy assumption (A1), inferring the treatment effect with said instruments can be highly misleading; see \citet{staiger_instrumental_1997}, \citet{stock_survey_2002}, \citet{dufour_identification_2003}, and \citet{andrews2005inference} for overviews on weak instruments. As such, it is common to conduct a ``pre-test'' to assess the strength of the instruments and only proceed with estimating and inferring the treatment effect when they are sufficiently strong. For example, \citet{swanson2013commentary} found that 93\% of IV papers that they surveyed in medical journals (or 75 out of 81 papers in their survey) conducted pre-tests for IV strength. The most popular pre-test is the F-test, which is based on computing the overall F-statistic from an ordinary least squares regression with the instruments as independent variables and the treatment as a dependent variable. If the F-test is above a certain threshold, usually $10$ \citep{staiger_instrumental_1997,stock_testing_2002}, assumption (A1) is considered to be satisfied and the investigator proceeds to test $H_0$. For other tests of instrument strength, see \citet{cragg_testing_1993}, \citet{stock_testing_2002}, and \citet{hahn_new_2002}.

Unfortunately, when it comes time to test the treatment effect, many IV analyses do not account for the pre-test done a priori. Roughly speaking, investigators compute p-values for $\beta^*$ based on 
 \begin{equation} \label{eq:2}
P_{H_0} (T \geq t), \quad{} t \in \reals
\end{equation}
where the distribution of $T$ is usually asymptotically Normal. In particular, the computation of the null distribution of $T$ in \eqref{eq:2} doesn't recognize that the data was used twice, first to assess instrument strength and second to infer the treatment effect; the null remains the same regardless of whether an F-test was done or not a priori, or whether the value of the F-statistic is 15 or 50. To put it differently,  \eqref{eq:2} neglects the fact that only the strongest, most promising instruments that passed the F-test are used to test for the treatment effect. Consequently, the resulting p-value and confidence interval (CI) of the treatment effect based on \eqref{eq:2} are biased and often overly optimistic (e.g. small p-values or shorter CIs). This problem also persists even if one uses weak-instrument robust tests as $T$, such as the Anderson-Rubin test \citep{anderson_estimation_1949} or the conditional likelihood ratio (CLR) test  \citep{moreira_conditional_2003} after determining that the instruments are weak from a pre-test. 

As a concrete example, in the work by \citet{card1993using} which we replicate in Section \ref{sec:app_geo}, the author used proximity to college as an instrument to estimate the effect of completed years of schooling on earnings. Card tested the effect of schooling on earnings by using two-stage least squares (TSLS), a popular method in IV to estimate treatment effect, and found a positive and significant effect (p-value: $0.016$). Unfortunately, Card's analysis of the treatment effect fails to take into account that an F-test for instrument strength was done a priori, with an observed F-statistic around 14; see page 11 and first row of Table 3 in \citet{card1993using}. Using our proposed method, after seeing that the F-test exceeded the typical threshold of 10, the effect of schooling on earnings with the TSLS test statistic goes away (p-value: $0.602$). More generally, in Section \ref{sec:simulation}, we find that if the F-test is near $10$, traditional approaches to inference based on \eqref{eq:2} inflate Type I error and has lower-than-expected coverage rates for confidence intervals, while our method that accounts for pre-testing controls the Type I error and covers $\beta^*$ at the nominal level.

\subsection{Why Not Sample Splitting?} \label{sec:sample_split}
A simple strategy of testing the treatment effect that accounts for instrument strength pre-testing is sample splitting. Specifically, a random subset of the data ($\Lambda_1$) is used for the pre-test and the rest ($\Lambda_2$) is used for testing $H_0$. In effect, this simplifies equation \eqref{eq:1} to
\[
P_{H_0} (T(\Lambda_2) \geq t \mid \text{tested IV strength with $\Lambda_1$}) = P_{H_0}(T(\Lambda_2) \geq t)
\] 
under the usual assumption of independence between $\Lambda_1$ and $\Lambda_2$. Then, the classic null distribution of $T$ in \eqref{eq:2} equals \eqref{eq:1}. However, a major downside of sample splitting is a loss of power, especially in small samples, to detect the treatment effect since only a subset of the sample is used to test $H_0$. Indeed, \citet{fithian_optimal_inference_2014} proved under general conditions that sample splitting is inadmissible to another procedure called {\emph{data carving}}, which uses the entire data for testing the treatment effect; they also showed that data carving with holdout significantly boosts statistical power compared to no holdout. Our proposed method uses a randomized pre-testing algorithm and is mathematically equivalent to data carving with holdout. In other words, our method uses all the data for both pre-testing and testing the treatment effect to achieve better power than sample splitting.

\subsection{Prior Work and Our Contributions}
While there are many works in weak instruments (see  \citet{stock_survey_2002} and \citet{andrews2005inference} for a review), only a handful of work have looked into the issue of inferring treatment effects after pre-testing. The most relevant among the few is by \citet{moreira2009tests}, who derived a quantity akin to \eqref{eq:1} when the test statistic $T$ is similar and the pre-test is a function of sufficient statistics. \citet{stock_testing_2002} also discussed various pre-tests for instrument strength, but their emphasis was on choosing different critical values for these pre-tests. Finally, \citet{andrews2005inference} highlighted the effect of pre-testing on Type I error of testing $H_0$.

We take a different approach than \citet{moreira2009tests} and derive \eqref{eq:1} by using recent advances in selective inference \citep{fithian_optimal_inference_2014, tian_randomized_2016, tian_selective_sampling_2016, lee_exact_lasso_2016, bi_inferactive_2017}. Broadly speaking, selective inference derives conditional distributions of test statistics for complex conditioning events such as those resulting from marginal screening \citep{lee_marginal_screening_2014}, forward stepwise regression \citep{taylor_forward_stepwise_2014, loftus_significance_2014-1}, and the Lasso \citep{tibshirani_regression_1996, lee_exact_lasso_2016}. The key technical steps in our method involve (i) rewriting the F-test as a solution to a randomized convex optimization problem and (ii) reparametrizing the conditional null in \eqref{eq:1} so that we can efficiently generate the null distribution using standard MCMC or hit-and-run sampling methods \citep{tian_randomized_2016, tian_selective_sampling_2016}. Our method is general in that it can accommodate popular test statistics for treatment effects, including the Wald test derived from the TSLS estimator, the Anderson-Rubin test, and the conditional likelihood ratio (CLR) test, where the latter is shown to be robust to weak instruments \citep{moreira_conditional_2003} and nearly optimal \citep{andrews_optimal_2006}. Also, while we focus on the popular F-test as our pre-test, one advantage of our method versus \citet{moreira2009tests} is that it can be modified to handle a wide variety of pre-tests, such as (i) a pre-test based on checking whether the estimated signs of the regression coefficients between the instruments and the treatment are congruent to what's expected from subject-matter knowledge \citep{andrews2017unbiased}, (ii) a pre-test based on applying a Lasso-type optimization to select the strongest instruments \citep{belloni2012sparse}, and (iii) a pre-test where the instruments are selected based on forward stepwise regression; see Section \ref{sec:extensions} for details. 

We show that our proposed method controls the conditional Type I error at level $\alpha$,
\begin{equation} \label{eq:typeI}
P_{H_0} (\text{reject } H_0 \mid \text{tested IV strength }) \leq \alpha
\end{equation}
This is also known as the selective type I error \citep{fithian_optimal_inference_2014}. By controlling \eqref{eq:typeI} at level $\alpha$, we can invert the test to achieve nominal $1-\alpha$ coverage rate of conditional/selective confidence intervals that condition on having tested for IV strength. We also remark that by controlling \eqref{eq:typeI} at level $\alpha$, we automatically control the usual marginal Type I error rate arising from \eqref{eq:2} and achieve $1-\alpha$ marginal coverage rates. By comparison, the usual confidence intervals without adjusting for the pre-test may fall below the $1-\alpha$ nominal rate.  We also demonstrate that when we apply our method with the aforementioned data from \citet{card1993using}, accounting for the pre-test leads to different conclusions about the treatment effect than not accounting for it.

\section{IV Model and Problem Statement} 

\subsection{Notation}
For each individual $i=1,2,\ldots, n$, let $Y_i \in \mathbb{R}$ be the outcome, $D_i \in \mathbb{R}$ be the treatment, and $\mathbf{Z}_i \in \mathbb{R}^p$ be the $p$ instruments. Denote the vector forms of these variables as $\mathbf{Y} = (Y_1, \ldots, Y_n)$ and $\mathbf{D} = (D_1, \ldots, D_n)$. Denote $Z$ to be the $n\times p$ matrix of instruments where we assume the matrix $Z$ is full rank. Let $I$ be the identity matrix. Define the projection matrix onto the column space of $Z$ as $P_Z = Z (Z^T Z)^{-1} Z^T$ and denote $P_{Z^{\perp}} = I - P_Z$ to be the residual projection matrix. Let $\mathbb{I}(\cdot)$ be the indicator function and $||\cdot ||_2$ be the usual $\ell_2$ norm of a vector. Let $\phi(\mu,\sigma^2)$ denote the density function of the Normal distribution with mean $\mu$ and variance $\sigma^2$. 

We adopt the usual big-O notation $O_p (\cdot)$ to denote the stochastic order of a function, i.e. $X_n = O_p (a_n)$ means that for any $\epsilon > 0$, there exists finite $M, N > 0$ such that $P(|X_n / a_n| > M) < \epsilon, \forall n > N$.

\subsection{Review: Model} \label{sec:model}
For the observed $Y_{i}, D_{i}$, and $\mathbf{Z}_{i}$, we consider a popular linear model in the IV literature that is also common in studying weak IVs \citep{andrews_optimal_2006,wooldridge_econometrics_2010}:
\begin{equation} 
\begin{aligned} \label{eq:model}
Y_i &=  D_i \beta^* + \delta_i \\
D_i &=  \mathbf{Z}_{i}^T \gamma^* + \xi_{i2} \\
(\delta_i, \xi_{i2} | \mathbf{Z}_i) &\iid N (0, \Sigma^*), \quad{} \mathbf{Z}_{i}\iid F
\end{aligned}
\end{equation}
Here, $\beta^*, \bm{\gamma}^*, \Sigma^*$ are unknown model parameters. The target parameter for inference is $\beta^*$, which we refer to as the treatment effect. The parameter $\Sigma^*$ represents the variance of the structural errors $\delta_i$ and $\xi_{i2}$ and $\Sigma$ does not depend on $Z$.  The parameter $\bm{\gamma}^*$ represents instruments' strength, with a large $\bm{\gamma}^*$ roughly translating to strong instruments and a small $\bm{\gamma}^*$ translating to weak instruments; note that assumption (A1) is satisfied so long as $\bm{\gamma}^* \neq \mathbf{0}$. Also, assumptions (A2) and (A3) are satisfied because the conditional mean of the structural errors $\delta_i$ and $\xi_{i2}$ given $Z_i$ is mean zero. Finally, we assume $\mathbf{Z}_i$ is randomly generated from a distribution $F$ where $E(\mathbf{Z}_i)$, $E(\mathbf{Z}_i \mathbf{Z}_i^T)$ exist and $E(\mathbf{Z}_i \mathbf{Z}_i^T)$ is positive definite. 

Without loss of generality, we assume that $Y, D, Z$ are centered to mean zero, which allows us to remove the intercept terms in equation \eqref{eq:model} by the Frisch-Waugh-Lovell Theorem \citep{davidson_estimation_1993}. By the same argument, we can also incorporate pre-instrument exogenous covariates $X$ and if the concatenated matrix of $Z$ and $X$ are full rank, we can residualize $X$ out from the concatenated matrix by the Frisch-Waugh-Lovell Theorem to arrive at model \eqref{eq:model}. Also, while we assume Gaussian errors similar to some work in weak IVs \citep{andrews_optimal_2006, andrews2007performance} to simplify the expression for the conditional null distribution in \eqref{eq:2}, the assumption can be removed so long as the test statistic for testing $\beta^*$ is asymptotically Normal; see Section \ref{sec:asymp_cond} for more details.

It is useful to write model \eqref{eq:model} into a reduced-form model where $Y$ and $D$ are only functions of $Z$: 
\begin{equation} \label{eq:reduced_model}
\begin{aligned}
Y_i &=  \mathbf{Z}_i^T (\bm{\gamma}^* \beta^*) + \xi_{i1} \\
D_i &=  \mathbf{Z}_{i}^T \bm{\gamma}^* + \xi_{i2} \\
(\xi_{i1}, \xi_{i2}) &\iid N (0, \Omega^*), \quad{} \xi_{i1} = \delta_i + \xi_{i2} \beta^*
\end{aligned}
\end{equation}
Note that for any value of $\beta^*$, there is an invertible mapping between the covariance matrix of the reduced-form model $\Omega^*$ and the covariance matrix of the structural model $\Sigma^*$, i.e.
\[
\Omega^* = \begin{pmatrix} 1 & \beta^* \\ 0 & 1 \end{pmatrix} \Sigma^* \begin{pmatrix} 1 & 0 \\ \beta^* & 1 \end{pmatrix} = \begin{pmatrix}\Omega_{11}^* & \Omega_{12}^* \\ \Omega_{21}^* & \Omega_{22}^* \end{pmatrix}
\]

\subsection{Review: Point Estimators for Model Parameters} \label{sec:point_est} 
We briefly review relevant point estimators for the model parameters in \eqref{eq:model}. First, for estimating the target parameter $\beta^*$, a popular and consistent estimator is the two stage least squares (TSLS) estimator, 
\begin{equation} \label{eq:tsls_est} 
\widehat{\beta}_{\rm TSLS} = \frac{\mathbf{D}^T P_Z \mathbf{Y}}{\mathbf{D}^T P_Z \mathbf{D}}
\end{equation}
Under fairly general conditions, of which our model \eqref{eq:model} satisfies, $\widehat{\beta}_{\rm TSLS}$ converges to $\beta^*$ in probability; see the supplementary materials or Section 5.2.1 of \citet{wooldridge_econometrics_2010} for details. TSLS derives its name because it can be constructed based on running ordinary least squares (OLS) regression twice. Specifically, in the first stage, the treatment $\mathbf{D}$ is regressed on the instruments $Z$ to obtain predicted values of $\mathbf{D}$, denoted as $\widehat{\mathbf{D}}$. In the second stage, the outcome $\mathbf{Y}$ is regressed on the predicted values of $\widehat{\mathbf{D}}$ and the estimated OLS coefficient for $\widehat{\mathbf{D}}$ is equal to $\widehat{\beta}_{\rm TSLS}$.

A popular, consistent estimator of the covariance matrix $\Omega^*$ in \eqref{eq:reduced_model} is 
\begin{equation} \label{eq:omega_est}
\widehat{\Omega} = \frac{1}{n-p} \begin{pmatrix}\mathbf{Y}^T \\ \mathbf{D}^T \end{pmatrix} P_{Z^\perp} \begin{pmatrix} \mathbf{Y} & \mathbf{D} \end{pmatrix}
\end{equation}
Again, under general conditions, including conditions typical in weak IV settings (c.f. Section 2.2.2 in \citet{andrews2007performance}), the above estimator $\widehat{\Omega}$ converges to $\Omega^*$ in probability. Then, under $H_0: \beta^* = \beta_0$, we can consistently estimate $\Sigma^*$ by using the mapping between the covariance matrices of $\Omega^*$ and $\Sigma^*$
\begin{align*} %\label{eq:varest}
\widehat{\Sigma}(\beta_0) &=  \frac{1}{n-p} \begin{pmatrix} \mathbf{Y}^T - \mathbf{D}^T \beta_{0}  \\ \mathbf{D}^T \end{pmatrix}  P_{Z^\perp} \begin{pmatrix} \mathbf{Y} - \mathbf{D}\beta_{0} & \mathbf{D} \end{pmatrix}
%\widehat{\Omega}(\widehat{\beta}_{\rm TSLS}) &= \begin{pmatrix} 1 & \widehat{\beta}_{\rm TSLS} \\ 0 & 1 \end{pmatrix} \widehat{\Sigma} \begin{pmatrix} 1 & 0 \\ \widehat{\beta}_{\rm TSLS} & 1 \end{pmatrix} \rightarrow \Omega^*
\end{align*}
where $\widehat{\Sigma}(\beta_0)$ converges to $\Sigma^*$ in probability under $H_0$. Alternatively, we can plug in a consistent estimator of $\beta^*$ in lieu of $\beta_0$, say the TSLS estimator in \eqref{eq:tsls_est}, to obtain a consistent estimator of $\Sigma^*$.

\subsection{Review: Tests for $\beta^*$} \label{sec:test}
We briefly review three tests for the null hypothesis of treatment effect $H_0: \beta^* = \beta_0$; see \citet{andrews2005inference}, \citet{wooldridge_econometrics_2010}, or the supplementary materials for additional details. The first test is based on the TSLS estimator of $\beta^*$ in \eqref{eq:tsls_est} and is denoted as $T_{\rm TSLS}(\beta_0)$,
\begin{equation*} \label{eq:tsls_test}
T_{\rm TSLS}(\beta_0) = \frac{\mathbf{D}^T P_Z (\mathbf{Y} - \mathbf{D} \beta_0)}{ \sqrt{\widehat{\Sigma}_{11}} \sqrt{\mathbf{D}^T P_Z \mathbf{D}}}
\end{equation*}
Here, $\widehat{\Sigma}_{11}$ is a consistent estimator of $\Sigma_{11}^*$. The test statistic \eqref{eq:tsls_test} is often called a Wald test because it can also be written in the ``Wald'' form $\sqrt{n}(\widehat{\beta}_{\rm TSLS} - \beta_0)$. Under $H_0$ and the assumptions underlying consistency of $\widehat{\beta}_{\rm TSLS}$, $T_{TSLS}(\beta_0)$ converges to a standard normal.  However, under weak instrument asymptotics of \citet{staiger_instrumental_1997}, $T_{\rm TSLS}(\beta_0) $ converges to a non-Normal distribution.

The second test statistic is the Anderson-Rubin (AR) test \citep{anderson_estimation_1949} and is denoted as $T_{\rm AR}(\beta_0)$ 
\begin{align*}
T_{\rm AR}(\beta_0) &= \frac{(\mathbf{Y} - \mathbf{D} \beta_0)^T P_Z (\mathbf{Y} - \mathbf{D} \beta_0) / p}{(\mathbf{Y} - \mathbf{D} \beta_0)^T (I - P_Z) (\mathbf{Y} - \mathbf{D} \beta_0) / (n-p)}
\end{align*}
Under $H_0$, the AR test follows an F distribution with degrees of freedom $p$ and $n-p$. Unlike the test statistic based on TSLS, the AR test still follows an F distribution under weak instrument asymptotics, or, more generally, for any value of $\gamma^*$  \citep{staiger_instrumental_1997}.

The third test is the conditional likelihood ratio (CLR) test \citep{moreira_conditional_2003}, which is based on exploiting invariance and sufficiency in model \eqref{eq:reduced_model}. Specifically, under a known covariance $\Omega^*$, Gaussian errors and fixed $Z$, the sufficient statistics of model \eqref{eq:reduced_model} are (c.f. section 2.2 of \cite{andrews2007performance})
\begin{align*}
\mathbf{U} &= \frac{(Z^T Z)^{-\frac{1}{2}} \cdot \widetilde{Y} \cdot \mathbf{b}_0}{\sqrt{\mathbf{b}_0^T \Omega \mathbf{b}_0}}, \quad{}
\mathbf{R} = \frac{(Z^T Z)^{-\frac{1}{2}}\cdot \widetilde{Y} \cdot \Omega^{-1} \cdot \mathbf{a}_0}{\sqrt{\mathbf{a}_0^T \Omega^{-1} \mathbf{a}_0}}, \quad{} \mathbf{a}_0 = \begin{pmatrix}
\beta_0 \\
1
\end{pmatrix}, \quad{}
 \mathbf{b}_0 = \begin{pmatrix}
1 \\ 
-\beta_0
\end{pmatrix}, \quad{}
\widetilde{Y} = \begin{pmatrix} Z^T \mathbf{Y} & Z^T \mathbf{D} \end{pmatrix}
\end{align*}
where  $\mathbf{U}$ and $\mathbf{R}$ are $p$ dimensional vectors. Let $LR(\beta_0)$ be the statistic that is a function of $\mathbf{U}$, $\mathbf{R}$, and $\beta_0$,
\begin{equation} \label{eq:clr}
LR (\beta_0) = \frac{1}{2}\left\{ Q_U - Q_R + \sqrt{ (Q_U + Q_R)^2 - 4(Q_U Q_R - Q_{UR}^2)} \right\}, \quad{}
Q = \begin{pmatrix} \mathbf{U}^T \mathbf{U} & \mathbf{U}^T \mathbf{R} \\ \mathbf{R}^T \mathbf{U} & \mathbf{R}^T \mathbf{R} \end{pmatrix}
\end{equation}
For notational convenience, we let $Q_U = \mathbf{U}^T \mathbf{U},\ Q_R = \mathbf{R}^T \mathbf{R}$, and $\ Q_{UR} = \mathbf{U}^T \mathbf{R}$. The CLR test is based on the null distribution of $LR(\beta_0)$ conditional on $Q_R$. \citet{moreira_conditional_2003} showed that the CLR test is robust against weak instruments.
\citet{andrews_optimal_2006} derived the analytical formula of this conditional null density with fixed $Z$, known $\Omega^*$, and Gaussian errors and showed that the test is a nearly uniformly most powerful test among similar invariant tests. Also,  they considered the null distribution of the CLR test under a random $Z$, unknown $\Omega^*$, and non-Gaussian errors with a plug-in estimate of $\Omega^*$ from \eqref{eq:omega_est}. More concretely, let
\begin{align*}
\widehat{U} &= \frac{(Z^T Z)^{-\frac{1}{2}} \cdot \widetilde{Y} \cdot \mathbf{b}_0}{\sqrt{\mathbf{b}_0^T \widehat{\Omega} \mathbf{b}_0}}, \quad{}
\widehat{R} = \frac{(Z^T Z)^{-\frac{1}{2}}\cdot \widetilde{Y} \cdot \widehat{\Omega}^{-1} \cdot \mathbf{a}_0}{\sqrt{\mathbf{a}_0^T \widehat{\Omega}^{-1} \mathbf{a}_0}}, \quad \widehat{Q} = \begin{pmatrix}\widehat{\mathbf{U}}^T \widehat{\mathbf{U}} & \widehat{\mathbf{U}}^T \widehat{\mathbf{R}} \\ \widehat{\mathbf{R}}^T \widehat{\mathbf{U}} & \widehat{\mathbf{R}}^T \widehat{\mathbf{R}}\end{pmatrix} = \begin{pmatrix} \widehat{Q}_{U} & \widehat{Q}_{UR} \\ \widehat{Q}_{UR} & \widehat{Q}_{R} \end{pmatrix}
\end{align*}
Then $\widehat{\rm LR}(\beta_0)$ is the CLR test statistic in \eqref{eq:clr} with a plug-in $\widehat{\Omega}$, i.e.
\begin{equation} \label{eq:clr_hat}
\widehat{\rm LR}(\beta_0) = \frac{1}{2} \left \{ \widehat{Q}_U - \widehat{Q}_R + \sqrt{ (\widehat{Q}_U + \widehat{Q}_R)^2 - 4 (\widehat{Q}_U \widehat{Q}_R - \widehat{Q}_{UR}^2)}\right\}
\end{equation}
Under weak instrument asymptotics and some regularity assumptions on $Q$, they showed that $\widehat{\rm LR}(\beta_0)$ is asymptotically equivalent to the null distribution of ${\rm LR}(\beta_0)$ with fixed $Z$, known $\Omega^*$, and Gaussian errors. Section 4.1 of \citet{andrews2007performance} and Section 3.1 of \citet{mikusheva2010robust} proposed fast algorithms to compute p-values and confidence intervals from the CLR test. Our derivation in section \ref{sec:weakiv} builds upon these theoretical results; for completeness, we reiterates the technical details in the supplementary materials.

\subsection{Review: Pre-Testing Instrument Strength with the F-test}
We review the most popular pre-test for assessing instrument strength,  the F-test. Formally, let $\widehat{\bm{\gamma}} = (Z^T Z)^{-1} Z^T \mathbf{D}$ be the OLS estimate of $\bm{\gamma}^*$ in equation \eqref{eq:model} and let $\bar{D}=n^{-1} \sum_{i=1}^n D_i$ be the average of $D_i$. Then, the F-test to assess instrument strength is the usual F-test from regressing $D_i$ on $\mathbf{Z}_{i}$ and testing $H_0: \bm{\gamma}^* = 0$ versus  $H_1: \bm{\gamma}^* \neq 0$:
\begin{equation} \label{eq:F} 
F = \frac{\frac{1}{p} \sum_{i=1}^n (\bar{D} - \mathbf{Z}_i^T \widehat{\bm{\gamma}})^2}{\frac{1}{n-p} \sum_{i=1}^n (D_i - \mathbf{Z}_i^T \widehat{\bm{\gamma}})^2} = \frac{\frac{1}{p} \sum_{i=1}^n (\mathbf{Z}_i^T \widehat{\gamma})^2}{\frac{1}{n-p} \sum_{i=1}^n (D_i - \mathbf{Z}_i^T \widehat{\bm{\gamma}})^2}
\end{equation}
where the second equality comes from centering $\mathbf{D}$ to mean zero. Typically, if the observed $F$ test for instrument strength exceeds a certain threshold $C_0$, usually $F \geq C_0 = 10$, the analyst would declare that the instruments satisfy (A1) and proceed with inferring the treatment effect $\beta^*$, most often with the TSLS test statistic; see, for example, the empirical example from \citet{card1993using} in Section \ref{sec:app_geo}. If the observed $F$ test is below the threshold so that the instruments are weak, the analyst may elect to test $\beta^*$ with weak-instrument robust tests such as the AR test or the CLR test \citep{stock_testing_2002,moreira2009tests}. Regardless of the choice of the test statistic for $\beta^*$, the F-test is common in practice to assess instrument strength.

\subsection{Problem Statement} \label{sec:prob_statement}
We now formally state the goal sketched out in equation \eqref{eq:1}. For a given test statistic $T$, the goal of the paper is to derive its null distribution conditional on the F-statistic in \eqref{eq:F}. In particular, for test statistics that perform well with strong instruments, such as the TSLS test statistic, we derive the null of $T$ given that the F-statistic exceeded the threshold $C_0$
\begin{equation} \label{eq:goal1}
P_{H_0} (T \geq t \mid F \geq C_0) 
\end{equation}
We also derive the null distribution of test statistics that are robust to weak instruments, such as the AR test or the CLR test, if the F-test is below the threshold. 
\begin{equation} \label{eq:goal2}
P_{H_0} (T \geq t \mid F < C_0) 
\end{equation}
The first conditional null \eqref{eq:goal1} mimics most investigators' analysis where they first test the instruments' strength and then proceed to test for $\beta^*$ only if the instruments passed the strength test. The second conditional null \eqref{eq:goal2} mimics a more cognizant investigator who uses a weak-instrument robust test after the instruments were deemed to be too weak from an F-test. Also, for the first conditional null \eqref{eq:goal1} where the instruments are deemed strong, we use traditional asymptotic arguments (i.e strong instruments asymptotics in Section 5.2 of \citet{wooldridge_econometrics_2010}). But, for the null in \eqref{eq:goal2}, because the instruments are deemed weak, we utilize weak instrument asymptotics of \citet{staiger_instrumental_1997} to obtain conditional null distributions.

We use the conditional nulls in equations \eqref{eq:goal1} and \eqref{eq:goal2} to construct tests that control the conditional Type I error at level $\alpha$ and to invert them to achieve conditional $1-\alpha$ coverage \citep{fithian_optimal_inference_2014,bi_inferactive_2017}. This is akin to using the traditional null distribution of tests  in \eqref{eq:2} to obtain p-values and confidence intervals. The difference between the traditional approach and our approach is that the conditional null recognizes that strength testing was done a priori and is arguable a more honest reflection of statistical uncertainty in a typical IV analysis. 

\section{Method} \label{sec:method}
We start with the derivation of the conditional null in \eqref{eq:goal1}, which is broken into three parts. First, we show that the conditioning event $\{F \geq C_0\}$ is asymptotically equivalent to a set of solutions from a penalized convex optimization program. Second, we show that the conditional null density of a test statistic $T$ is the product of the null density of $T$ without the conditioning event, say the standard Normal for the TSLS test statistic or the F distribution for the Anderson-Rubin test statistic, and a set of simple quadrant or box constraints arising from the aforementioned penalized convex program. Third, we apply standard MCMC algorithm such as Metropolis-Hasting or Hit-and-Run sampling to sample from the conditional null density. Section \ref{sec:weakiv} extends the result in Section \ref{sec:method} by (i) deriving the conditional null in \eqref{eq:goal2} under weak instrument asymptotic and (ii) discussing other pre-tests that can be used in our method.

\subsection{Equivalence Between the F-test and a Penalized Convex Algorithm}
The first step in our method is to establish that the conditioning event $\{F \geq C_0\}$ is asymptotically equivalent to a set of solutions from a penalized convex optimization problem. This equivalence allows us to use the rich set of methods developed in selective inference and derive conditional densities where the conditioning events are expressed as solutions to convex optimization problems \citep{tian_selective_sampling_2016}.

\begin{lemma}  \label{lem:optim}  
Let $\mathbf{S} = (Z^T Z)^{-\frac{1}{2}} Z^T \mathbf{D} \in \reals^p$. For any $C_0 \in \reals$ in the conditioning event $\{F \geq C_0\}$, define the following penalized convex optimization program
\begin{equation} \label{eq:optim_v}
\widehat{\mathbf{v}} = \amin{\mathbf{v}} \frac{1}{2} ||\mathbf{v} - \mathbf{S}||^2_2 + \lambda ||\mathbf{v}||_2, \quad{} \lambda = \sqrt{C_0 \frac{p}{n-p} \left(\sum_{i=1}^n (D_i - \mathbf{Z}_i^T \widehat{\bm{\gamma}})^2\right)}
\end{equation}
Then, as $n \to \infty$, we have the following asymptotic equivalence between the events
\[
\mathbb{I} (F \geq C_0) = \mathbb{I} (\widehat{\mathbf{v}} \neq 0)  + O_p \left(\frac{1}{\sqrt{n}} \right)
\]
\end{lemma}
An immediate consequence of Lemma \ref{lem:optim} is that it allows us to focus on the density of the null distribution of $T$ conditional on $\widehat{\mathbf{v}} \neq 0$ instead of being conditional on $F \geq C_0$, i.e.
\begin{equation} \label{eq:select_dist}
\ell_{\beta_0}(T\mid F\geq C_0) \approx \ell_{\beta_0} (T \mid \widehat{\mathbf{v}} \neq 0)
\end{equation}
where $\ell(\cdot)$ denotes a density function of random variables. The intuition behind Lemma \ref{lem:optim} follows from considering an idealized $F$ test where the denominator of the $F$ statistic in \eqref{eq:F} is assumed to be known. Then, the numerator of the $F$ test is a quadratic function of $Z^T \mathbf{D}$ and $\mathbf{D}^T Z^T (Z^T Z)^{-1} Z^T \mathbf{D}$, giving us the ``loss'' function in equation \eqref{eq:optim_v}. The final step involves choosing the right penalty $\lambda$ to achieve asymptotic equivalence. Interestingly, the optimization problem in \eqref{eq:optim_v} is a special case of the group Lasso with one single group \citep{tian_selective_sampling_2016}. 

\subsection{Improving Power and Computation: A Randomized F-test with a Directional Constraint} \label{sec:rand_F}
While Lemma \ref{lem:optim} provides an equivalence relationship between tests and optimization problems in order for us to use the rich set of tools in selective inference, there are two caveats that need to be addressed for the Lemma to be practically useful. First, previous results in selective inference have shown that simply considering the original optimization problem like \eqref{eq:optim_v} may have less power for downstream inference than a randomized version of the original optimization problem \citep{fithian_optimal_inference_2014,  tian_randomized_2016}. Second, the conditioning event $\widehat{\mathbf{v}} \neq 0$ in Lemma \ref{lem:optim} is numerically unstable, especially if $\widehat{\mathbf{v}}$ is very close to zero, and may cause computational issues. This subsection addresses these two concerns.

To address the first concern, consider a randomized version of the convex program \eqref{eq:optim_v}
\begin{equation} \label{eq:randmodel}
\widehat{\mathbf{v}} = \amin{\mathbf{v}} \frac{1}{2} ||\mathbf{v} - \mathbf{S}||_2^2 + \lambda ||\mathbf{v}||_2 - \bm{\omega}^T \mathbf{v}
\end{equation}
where $\bm{\omega}$ is random noise from a known distribution $g(\bm{\omega})$ that is specified by the analyst and independent of $\mathbf{Y}$ and $\mathbf{D}$. For simplicity, we consider $g(\cdot)$ to be a Normal distribution in this paper, but our method works for non-Normal distributions. The randomized quadratic problem in \eqref{eq:randmodel} is strongly convex and a solution always exists. Also, while counterintuitive, \citet{fithian_optimal_inference_2014} and \citet{ tian_randomized_2016} showed that inference based on adding noise to the conditioning event via \eqref{eq:randmodel} can drastically improve power of tests with minimal changes to the conditioning event, especially for sufficiently small $\bm{\omega}$. Critically, \citet{tian_randomized_2016} showed that a randomized algorithm like \eqref{eq:randmodel} is mathematically equivalent to data carving with holdout \citep{fithian_optimal_inference_2014}, which as mentioned in Section \ref{sec:sample_split}, dominates sample splitting. % note that sample splitting is also a randomized algorithm. 

Also, at a high level, injecting a randomized term $\bm{\omega}$ has the additional benefit of allowing practitioners to directly use pre-existing test statistics for the treatment effect in IV instead of developing a new test statistic that has a tractable conditional null distribution. To better understand this, note that sample splitting, a type of randomized algorithm, allows practitioners to use any pre-existing test statistics to characterize their conditional null in Section \ref{sec:sample_split}. Injecting the randomized term $\bm{\omega}$ in \eqref{eq:optim_v} retains the same benefit as sample splitting in that practitioners can use any pre-existing test statistics, but is provably more powerful than sample splitting; see \citet{tian_randomized_2016} for additional discussions.

To address the second concern, let $\mathbf{u} = \widehat{\mathbf{v}} / ||\widehat{\mathbf{v}}||_2$ so that $\|\mathbf{u}\|_2 = 1$ and write $\mathbf{v} = d \cdot \mathbf{u}, \ d > 0$. Suppose we define a finer conditioning event than \eqref{eq:select_dist} where we condition on both $d$ and $\mathbf{u}$,
\begin{equation} \label{eq:select_dist_u}
\ell_{\beta_0} (T \mid d>0, \mathbf{u})
\end{equation}
The conditional density in \eqref{eq:select_dist_u} is essentially conditioning on the active direction of $\widehat{\mathbf{v}}$. Also, controlling \eqref{eq:select_dist_u} will still control the conditional Type I error of \eqref{eq:select_dist} since we are conditioning on a finer filtration. However, unlike \eqref{eq:select_dist}, \eqref{eq:select_dist_u} removes the numerical instability associated with sampling near the region $||\widehat{\mathbf{v}}||_2 \approx 0$; see \citep{tian_selective_sampling_2016} for a related problem. As such, we can using any off-the-shelf MCMC algorithm, say Gibss, without modification; see Section \ref{sec:example_asymp_dists} for details.

The rest of the sections are devoted to deriving \eqref{eq:select_dist_u}. Also, for the interested reader, we also derive \eqref{eq:select_dist} if $\widehat{\mathbf{v}}$ is a solution to a randomized algorithm \eqref{eq:randmodel} without imposing the numerically stabilizing directional constraint in the supplementary materials.

\subsection{Exact Conditional Null Density} \label{sec:select_dists}
We begin the derivation of \eqref{eq:select_dist_u} by assuming $Z$ is fixed and the nuisance parameters $\bm{\gamma}^*,\Sigma^*$ are known. This allows us to derive exact, non-asymptotic conditional null densities using elementary probability theory; we later remove these simplifying assumptions by invoking asymptotic arguments.

We begin by stating the Karush-Kuhn-Tucker (KKT) condition that the solution $\mathbf{v}$ in \eqref{eq:randmodel} must satisfy.
$$
\bm{\omega} = \mathbf{v} - \mathbf{S} + \lambda \frac{\mathbf{v}}{||\mathbf{v}||_2} = d \mathbf{u} - \mathbf{S} + \lambda \mathbf{u}
$$
The KKT condition establishes a relationship between the randomization term $\bm{\omega}$ and the variables in the optimization problem, $\mathbf{v} = d \cdot \mathbf{u}$  and $\mathbf{S}$. In other words, the KKT conditions provides a mapping to reparametrize the conditional density from $(\mathbf{S},\bm{\omega})$ to $(\mathbf{S},d,\mathbf{u})$ with $ ||\mathbf{u}||_2 = 1$ by using a simple change-of-variables formula:
\begin{align} 
\ell_{\beta_0} (\mathbf{S}, \bm{\omega} \mid d > 0, \mathbf{u}) %&= f_{\beta_0} (S) \cdot g(\omega)\cdot \mathbb{I} (\widehat{v} \neq 0) \\ 
&= f_{\beta_0} (\mathbf{S}) \cdot g\left(d \mathbf{u} - \mathbf{S} + \lambda \mathbf{u}\right)\cdot \mathbb{I} (d > 0) \cdot |\mathcal{J}|, \quad{} |\mathcal{J}| = (d + \lambda)^{(p-1)}
\label{eq:tsls_asymp_density}
\end{align}
Here, $\mathcal{J}$ is the Jacobian from the change-of-variables formula (c.f. section 5.4 in \citet{tian_selective_sampling_2016}) and $f_{\beta_0}(\mathbf{S})$ is the original density of $\mathbf{S}$ under model \eqref{eq:model} and $H_0$. Broadly speaking, the terms on the right hand side of $f_{\beta_0}(\mathbf{S})$ in \eqref{eq:tsls_asymp_density} represent the effect that the F-test has on $\mathbf{S}$. These terms essentially reweigh and constrain $f_{\beta_0}(\mathbf{S})$ to reflect that the investigator has already tested IV strength and are based on the optimization variable $\mathbf{v}$. Also, because the F-test or any pre-test involving IV strength only affects the distribution of $\mathbf{D}$, the terms to the right of $f_{\beta_0}(\mathbf{S})$ in \eqref{eq:tsls_asymp_density} are only functions of $\mathbf{D}$ (via $\mathbf{S}$). 

Now, if the density $f_{\beta_0}(\mathbf{S})$ is known, i.e. if we know the model parameters $ \bm{\gamma}^*, \Sigma^*$ that govern the distribution of $\mathbf{D}$, and if the test statistic $T$ is a function of $\mathbf{Y}$ and $\mathbf{S}$ only, one can directly use \eqref{eq:tsls_asymp_density} in an MCMC sampler to generate samples of the $\mathbf{Y}$ and $\mathbf{S}$ under $H_0$ and then plug the MCMC samples into the test statistic $T$ to generate a finite-sample conditional null distribution of $T$. However, in practice, these model parameters are unknown and act as nuisance parameters. In the next section, we get rid of these nuisance parameters by deriving an asymptotic version of the conditional null density.

\subsection{Asymptotic Conditional Null Density} \label{sec:asymp_cond}
We start by providing a heuristic argument for our asymptotic analysis. Suppose the distribution of $(T,S)$ is asymptotically Normal without pre-testing, i.e.
\begin{equation} \label{eq:CLT}
\begin{pmatrix}T \\ \mathbf{S}\end{pmatrix} \rightarrow N \left\{\begin{pmatrix}\mu_T \\ \bm{\mu}_\mathbf{S}\end{pmatrix}, \begin{pmatrix} W_T & \mathbf{W}_{\mathbf{S},T} \\ \mathbf{W}_{\mathbf{S},T} & \mathbf{W}_S\end{pmatrix}\right\}
\end{equation}
where $\mathbf{W}_{S}, \mathbf{W}_{\mathbf{S},T}$, and $W_{T}$ are known. Let $\mathbf{O} = \mathbf{S} - \mathbf{W}_{\mathbf{S}, T} W^{-1}_{T} T$. Then, we can rewrite $\mathbf{S}$ as $\mathbf{S} = \mathbf{W}_{\mathbf{S}, T} W^{-1}_{T} T + \mathbf{O}$ and, by the joint Gaussianity of $(T,\mathbf{S})$, $\mathbf{W}_{\mathbf{S}, T} W^{-1}_{T} T$ and $\mathbf{O}$ are asymptotically independent of each other. Additionally, from the properties of multivariate Gaussians, conditional on $\mathbf{O}$, the distribution of $\mathbf{S}$ is no longer dependent on $\bm{\mu}_\mathbf{S}$ in the asymptotic sense; in other words, $\mathbf{O}$ is asymptotically sufficient for the distribution of  $\mathbf{S}$. %$\mu_T$ is known under the null and in many common cases $\mu_T = \beta^*$. 
Then, by applying a change-of-variables formula again, we can convert the distribution of $(T, \mathbf{S}, \bm{\omega})$ into $(T, \mathbf{O}, \bm{\omega})$ and condition on $\mathbf{O}$ to remove the dependence on $\bm{\mu}_\mathbf{S}$ associated with $f_{\beta_0}(\mathbf{S})$. Consequently, the tail probability of the test statistic $T$ conditional on $d>0, \mathbf{u}$ and $\mathbf{O}$, i.e.
\begin{equation} \label{eq:asymp_pivot}
\mathcal{P} (t) = P_{H_0} (T \geq t \mid d>0, \mathbf{u}, \mathbf{O})
\end{equation}
would be pivotal and asymptotically uniformly distributed under the null.

Formally, for a given sample size $n$, let $\mathcal{F}_n$ be the class of distributions of $(\mathbf{Y}, \mathbf{D}, Z)$ under \eqref{eq:model} and $\mathbb{F}_n \in \mathcal{F}_{n}$ be a sequence of null distributions of $(\mathbf{Y}, \mathbf{D}, Z)$ under $H_0$. Let $\mathcal{F}_{n}^c$ be the class of conditional distribution of $(\mathbf{Y}, \mathbf{D}, Z)$ given the selection event $\{d>0, \mathbf{u}, \mathbf{O}\}$ and let $\mathbb{F}_{n}^c \in \mathcal{F}_{n}^c$ be a sequence of conditional null distributions of $(\mathbf{Y}, \mathbf{D}, Z)$ under $H_0$. Let 
\[
\bm{\zeta}_i = \mathbf{Z}_i \begin{pmatrix} \delta_i \\ D_i \end{pmatrix}, \quad{}
\mathbb{E}_{\mathbb{F}_n} (\bm{\zeta}_i) = \bm{\mu}_n = \begin{Bmatrix}0 \\ \mathbb{E}_{\mathbb{F}_n} (\mathbf{Z}_i \cdot \mathbf{Z}_i^T) \bm{\gamma}^* \end{Bmatrix}
\]
where the expectation $\mathbb{E}_{\mathbb{F}_n}$ is under the distribution $\mathbb{F}_n$. Assumptions \ref{assump:1} and \ref{assump:2} below place different set of restrictions on the randomization density $g$ and the distribution of the data $\mathcal{F}_n$.
\begin{assumption} Suppose $\mathcal{F}_n$ and $g$ satisfy the following constraints. \label{assump:1}
\begin{enumerate}
\item $g$ is Normally distributed with mean $0$ and variance $c\cdot I_p$.
\item There exists $b>0$ and a constant $C_b$ such that $\underset{n\geq 1}{\sup}\ \underset{\mathbb{F}_n \in \mathcal{F}_n}{\sup} \mathbb{E}_{\mathbb{F}_n} \left(e^{b ||\bm{\zeta} - \bm{\mu}_n||_2^2}\right) \leq C_b$.
\end{enumerate}
\end{assumption}
\begin{assumption} Suppose $\mathcal{F}_n$ and $g$ satisfy the following constraints. \label{assump:2}
\begin{enumerate}
\item $g$ follows $g(\bm{\omega}) = \exp \{-\widetilde{g}(\bm{\omega})\} / C_g$ where $C_g$ is a normalizing constant and $\widetilde{g}$ has bounded derivatives up to the third order.
\item There exists $\tau >0$ and a constant $C_{\tau}$ such that $\underset{n\geq 1}{\sup}\ \underset{\mathbb{F}_n \in \mathcal{F}_n}{\sup} \mathbb{E}_{\mathbb{F}_n} \left( e^{\tau ||\bm{\zeta} - \bm{\mu}_n ||_1}\right) \leq C_{\tau}$.
\end{enumerate}
\end{assumption}
 Assumption \ref{assump:1} states that the randomization density $g$ is Gaussian and the distribution of the data is sub-Gaussian. Assumption \ref{assump:2} states that $g$ is a heavy-tail distribution and the moment generating function of $\mathbb{F}_n$ is uniformly bounded. Between the two assumptions, Assumption \ref{assump:1} makes a less stringent assumption on $\mathcal{F}_n$ compared to that from Assumption \ref{assump:2}, but at the expense of making a Normality assumption on $g$. Also, as mentioned before, $g$ can be chosen by the investigator and hence, the first parts of Assumptions \ref{assump:1} and \ref{assump:2} can be satisfied by design.

Lemma \ref{thm:linear} states that under either Assumption \ref{assump:1} or \ref{assump:2}, the p-values generated from the conditional null distribution in \eqref{eq:asymp_pivot} are uniformly distributed. 
\begin{lemma}[Asymptotic Validity of Conditional P-Values] \label{thm:linear} 
Suppose $(T, \mathbf{S})$ satisfy \eqref{eq:CLT} and $Z$ has up to sixth moments. Also, suppose that there are consistent estimators of $W_{T}, \mathbf{W}_{\mathbf{S},T}$, and $\mathbf{W}_{\mathbf{S}}$ as $n \to \infty$. Then, under either Assumptions \ref{assump:1}. or \ref{assump:2}, $\mathcal{P}(T)$ is asymptotically %uniformly 
pivotal, i.e. 
\begin{equation} \label{eq:cond_pivot}
\lim_{n\rightarrow \infty} \sup_{t \in [0,1]} \mid P_{\mathbb{F}_n^c}\{\mathcal{P} (T) \leq t \} - t \mid = 0
\end{equation}
\end{lemma}
Lemma \ref{thm:linear} states that p-value constructed from the conditional distribution in \eqref{eq:asymp_pivot} is uniform and the conditional Type I error can be asymptotically controlled. Specifically, by computing the p-value as the tail probability of a test statistic $T$ whose distribution is governed by \eqref{eq:asymp_pivot}, the probability that this p-value is less than $\alpha$ is asymptotically $\alpha$ under the null hypothesis. Also, $T$ in Lemma \ref{thm:linear} does not necessarily have to be a test statistic, per se; $T$ could be an asymptotically Normal estimator for $\beta^*$ (or more generally satisfy \eqref{eq:CLT}) and Lemma \ref{thm:linear} would provide a means to characterize the asymptotic distribution of $T$ conditional on passing the pre-test.  
Finally, Lemma \ref{thm:linear} fits into the more general result of Corollary 14 and Corollary 16 in \citet{markovic_bootstrap_2017} and we provide detailed connections in the supplement.  

Consistent estimators of variances $W_{T}, \mathbf{W}_{\mathbf{S},T}$, and $\mathbf{W}_\mathbf{S}$ in Lemma \ref{thm:linear} will depend on the type of test statistic for the treatment effect. Section \ref{sec:example_asymp_dists} shows that in many cases, the estimators can be found by replacing the covariances $\Sigma^*$ with a consistent estimator $\widehat{\Sigma}$. Alternatively, we can use the bootstrap to estimate the variances so long as they are consistent for $W_{T}$, $\mathbf{W}_{\mathbf{S},T}$, and $\mathbf{W}_{\mathbf{S}}$; see \citet{tian_randomized_2016} for details. 

\subsection{Example: TSLS After Testing Instrument Strength} \label{sec:example_asymp_dists}
This section details one use of Lemma \ref{thm:linear} to derive conditional null distribution of the TSLS test statistic $T_{\rm TSLS}$ after passing the F-test for instrument strength. The supplementary materials contain other examples of using Lemma \ref{thm:linear}, such as deriving the conditional null distribution of the the TSLS estimator $\widehat{\beta}$ after passing the F-test and the conditional null distribution of the Anderson-Rubin test after passing (or not passing) the F-test.
\begin{theorem}[Asymptotic Conditional P-Value of TSLS] \label{cr:tsls_cond} 
Consider the following conditional density of the TSLS test statistic $T_{\rm TSLS}$ conditional on passing the F-test, the active direction, and the sufficient statistic $O$ 
\begin{align} \label{eq:tsls_cond}
\ell_{\beta_0} (T_{\rm TSLS}, d \mid d>0,\mathbf{u},\mathbf{O}) &= C_{T} \cdot \phi_{[0,\widehat{W}_{T}]} (T_{\rm TSLS}) \cdot g\left\{-\widehat{\mathbf{W}}_{\mathbf{S},T} \widehat{W}_T^{-1} \cdot T_{\rm TSLS} + (d+ \lambda) \mathbf{u} - \mathbf{O}\right\} \cdot |\mathcal{J}| \cdot \mathbb{I}(d > 0) \\
\widehat{W}_T = 1, \quad{} \widehat{\mathbf{W}}_{\mathbf{S},T} &=\frac{ \widehat{\Sigma}_{12} (Z^T Z)^{-\frac{1}{2}}Z^T \mathbf{D}}{\sqrt{\widehat{\Sigma}_{11}}\sqrt{\mathbf{D}^T P_Z \mathbf{D}}} = \frac{\widehat{\Sigma}_{12} \mathbf{S}}{ \sqrt{\widehat{\Sigma}_{11} \mathbf{S}^T \mathbf{S}}} , \quad{} \mathbf{O} = \mathbf{S} - \widehat{\mathbf{W}}_{\mathbf{S},T} \widehat{W}_T^{-1} \cdot \widehat{T}_{\rm TSLS} \nonumber
\end{align}
where $C_{T}$ is a normalizing constant. Define $\widehat{\mathcal{P}}(t) = P_{\bar{\ell}_{\beta_0}} (T_{\rm TSLS} \geq t)$ where $\bar{\ell}_{\beta_0} = \int_{d'} \ell_{\beta_0} (T_{\rm TSLS}, d' \mid d'>0,\mathbf{u},\mathbf{O})$ is the marginal of $\ell_{\beta_0} (T_{\rm TSLS}, d \mid d>0,\mathbf{u},\mathbf{O})$. Then, under the assumptions in Lemma \ref{thm:linear}, we have 
\begin{equation*} \label{eq:asymp_p} 
\forall \delta>0, \quad \lim_{n\rightarrow \infty} P_{\mathbf{F}_{n}^c} \left\{ \sup_{t\in \mathbb{R}} |\widehat{\mathcal{P}}(t) - \mathcal{P}(t)| > \delta\right\} = 0 \text{ and } \lim_{n\rightarrow \infty} \sup_{t \in [0,1]} \left| P_{\mathbf{F}_{n}^c} \{ \widehat{\mathcal{P}}(T) \leq t \} - t \right| = 0
\end{equation*}
\end{theorem}
Unlike the exact conditional density in \eqref{eq:tsls_asymp_density}, the asymptotic conditional density of the TSLS test statistic in Theorem \ref{cr:tsls_cond} does not depend on unknown quantities like $f_{\beta_0}(\cdot)$. Specifically, for the TSLS test statistic, the conditional density of $T_{\rm TSLS}$ is essentially the standard Normal density, i.e. the usual distribution of $T_{\rm TSLS}$ under $H_0$ without pre-testing (i.e. $\phi_{[0,\widehat{W}_{T}]}({\rm TSLS})$ in \eqref{eq:tsls_cond}), reweighed by constraints from testing for instrument strength (i.e. everything to the right of $\phi_{[0,\widehat{W}_{T}]}({\rm TSLS})$ in \eqref{eq:tsls_cond}).

Given a conditional null density of a test statistic, we can use any off-the-shelf MCMC sampling method to sample from it and conduct hypothesis testing. For completeness, we detail one MCMC sampling algorithm, based on Gibbs, and use the conditional density of the TSLS test statistic in \eqref{eq:tsls_cond} as an example. Suppose we write the conditional density of the TSLS test statistic in \eqref{eq:tsls_cond} as follows
\begin{align*}
\ell_{\beta_0} (T_{\rm TSLS}, d \mid d>0,\mathbf{u},\mathbf{O}) &= C_T\cdot \phi_{[\beta_0,\widehat{W}_{T}]} (T_{\rm TSLS} ) \cdot g\left(-\widehat{\mathbf{W}}_{\mathbf{S},T} \widehat{W}_T^{-1} \cdot T_{\rm TSLS} + d \mathbf{u} + \lambda \mathbf{u} - \mathbf{O}\right) \cdot |\mathcal{J}| \cdot \mathbb{I}(d > 0) \\
&\propto \phi_{[\beta_0,\widehat{W}_{T}]} (T_{\rm TSLS}) \cdot g\left(-\widehat{\mathbf{W}}_{\mathbf{S},T} \widehat{W}_T^{-1} \cdot T_{\rm TSLS} + d \mathbf{u} + \lambda \mathbf{u} - \mathbf{O}\right) \cdot \mathbb{I}(d > 0) \\
&= h(T_{\rm TSLS},d)
\end{align*}
We will use $h(t,d)$ in the Gibbs sampler in Algorithm \ref{alg:sampling} to sample $t$.
\begin{algorithm}[tb!]
\caption{Gibbs Sampler for the Conditional Inference}
\label{alg:sampling}
\begin{algorithmic}
%\begin{enumerate}
\STATE {\textbf{Initiation}:} Set step $k= 0$. Initialize $(t, d)$ and denote them as $\{t^{(0)}, d^{(0)}\}$.
\STATE {\textbf{Gibbs Update}:} At step $k$ with variables $\{ t^{(k)}, d^{(k)} \}$
\begin{enumerate}
\item Update $t$ by sampling a proposal, denoted as $\tau^{(k+1)}$, from a proposal distribution that is Normal with mean $t^{(k)}$ and variance $a_k^2$;  $a_k$ represents the step size. The proposal distribution is symmetric with an acceptance ratio of 
$$
u = \min\left[ \frac{h\{t^{(k+1)}, d^{(k)}\}}{h\{t^{(k)}, d^{(k)}\}}, 1\right]
$$
Accept the proposal $t^{(k+1)}$ with probability $u$. Otherwise, set $t^{(k+1)} = t^{(k)}$.
\item Update $d$ by sampling a proposal, denoted as $d^{(k+1)}$, from a proposal distribution that is Normal with mean $d^{(k)}$ and variance $b_k^2$;  $b_k$ represents the step size. The proposal distribution is symmetric with an acceptance ratio of 
$$
u = \min \left[ \frac{h\{t^{(k+1)}, d^{(k+1)}\}}{h\{t^{(k+1)}, d^{(k)}\}}, 1 \right]
$$
Accept $d^{(k+1)}$ with probability $u$, otherwise $d^{(k+1)} \leftarrow d^{(k)}$.
\end{enumerate}
\STATE {\textbf{Stop}:} when convergence criterion is reached. Return samples for $t$ only.
%\end{enumerate}
\end{algorithmic}
\end{algorithm}

We take a moment to highlight some important implementation details in the Gibb sampling algorithm in \ref{alg:sampling}. First, we initialize $t^{(0)}$ to be the observed TSLS test statistic $\widehat{t}_{\rm TSLS}$ and $d^{(0)}$ to be the solution from the randomized convex program \eqref{eq:randmodel}. Second, if $g$ is Normal, a simple hit-and-run sampling algorithm can be used without the need for a step size; in this case, the sampling density reduces to a truncated gaussian with quadrant constraint on $d$; see \citet{bi_inferactive_2017} for details.

Once we obtain $N$ samples of $t_i, i=1,\ldots,N$ from the Gibbs sampling algorithm, we can use them to obtain conditional p-values and confidence intervals described in Section \ref{sec:prob_statement}. For example, let $\widehat{t}_{\rm TSLS}$ be the usual observed value of the test statistic and suppose we ran an F-test a priori to assess instrument strength where the F-statistic exceeded the value $C_0$. Instead of comparing $\widehat{t}_{\rm TSLS}$ to a marginal Normal null distribution to obtain its p-value, we can compare it against the samples $t_i$ generated from the Gibbs sampler and compute a conditional p-value $p(\beta_0)$ as follows.
\[
p(\beta_0) = \frac{1}{N} \sum_{i=1}^N \mathbb{I} (t_i \geq \widehat{t}_{TSLS} )
\]
If the significance level is set to $\alpha$, we can reject the null hypothesis of the treatment effect $H_0:\beta^* = \beta_0$ after observing that the instrument passed the F-test threshold of $C_0$ when $p(\beta_0) < \alpha$. More concretely, if $\beta_0 = 0$ and $p(0) < \alpha$, we can say that there is a significant treatment effect after \emph{accounting} for the instrument strength test done a priori. Also, by the duality between confidence intervals and hypothesis testing, we can construct a $1-\alpha$ conditional confidence interval of $\beta^*$ by retaining all $\beta_0$'s where $p(\beta_0) \geq \alpha$. 

\section{Extensions}
\subsection{Weak IV Asymptotics and the CLR Test}  \label{sec:weakiv}
Our asymptotic conditional nulls in Section \ref{sec:asymp_cond} relied on ``strong instrument'' asymptotics where as $n \to \infty$, the limiting distribution of test statistics without pre-testing was Normal. This was because the instrument passed the F-test for strength and was deemed to be sufficiently strong. However, when weak instruments are present and they fail to pass the F-test, weak-instrument robust test statistics are used to mitigate weak instrument biases and weak instrument asymptotics are used to study theoretical properties of test statistics \citep{staiger_instrumental_1997, andrews_optimal_2006, andrews2007performance}. In this section, we use weak IV asymptotics to derive the conditional null density of the CLR test, an almost-uniformly most powerful invariant (UMPI) test among similar tests \citep{andrews_optimal_2006}, when we fail to pass the F-test. For this section only, at the expense of power, we focus on the original non-randomized F-test defined in \eqref{lem:optim} as our pre-test to preserve the invariance property of the CLR test \citep{andrews_optimal_2006}. In particular, for the randomized F-test in \eqref{eq:optim_v}, if the instrument fails to pass the threshold $C_0$, the conditioning event $\{\widehat{\mathbf{v}} = 0\}$ is no longer a function of $Q$ which make up the CLR test statistic and the resulting conditional inference will not be invariant to the model parameters $(\beta, \lambda_{\bm{\gamma}})$ \citep{andrews_optimal_2006}.

Consider the following weak-IV asymptotic assumptions.
\begin{assumption} \label{assup:weak_iv} For some non-stochastic $p$ dimensional vector $\mathbf{C}_p \in \reals^p$, we have (i) $\bm{\gamma}^* = \mathbf{C}_{p} / \sqrt{n}$, (ii) $p$ does not depend on $n$, and (iii) $\beta^*$ is fixed for all $n$.
\end{assumption}
\noindent Part (i) of Assumption \ref{assup:weak_iv} states that the parameter $\bm{\gamma}^*$ that governs IV strength is local to zero. Part (ii) states that $p$ doesn't grow with $n$. Part (iii) of Assumption \ref{assup:weak_iv} states that the alternative $\beta$ is fixed and not local to the null value $\beta_0$ as $n$ grows. Parts (i)-(iii) combined make up what's commonly known as weak instrument asymptotics; see Section 6.1 of \citet{andrews2007performance} for additional discussions.

Under Assumption \ref{assup:weak_iv}, non-random  $Z$, and known $\Omega^*$, the asymptotic distribution of $Q$ follows a Wishart distribution \citep{andrews2007performance}. When $Z$ is random and $\Omega^*$ is unknown, let $q$ denote a specific value of $\widehat{Q}$, i.e.
\[
q = \begin{pmatrix}q_U & q_{UR} \\ q_{UR} & q_R\end{pmatrix} \quad q_U, q_R \in \mathbb{R}^{\texttt{+}}
\]

Theorem \ref{cr:Q_pval} derives the p-value of the CLR test after not passing the F-test.
\begin{theorem}[Asymptotic Conditional P-value of the CLR Test]  \label{cr:Q_pval}
%Let $\widehat{U}_2 = \widehat{Q}_{UR} / \sqrt{\widehat{Q}_U \widehat{Q}_R}$ 
Suppose Assumption \ref{assup:weak_iv} hold. Consider the plug-in CLR test statistic ${\widehat{\rm LR}}(\beta_0)$ in \eqref{eq:clr_hat} and let $n\to \infty$.
\begin{enumerate}
\item[(i)] Under $H_0$, the tail probability of CLR test statistic conditional on not passing the F-test is given by the integral below 
\begin{align*}
&\lim_{n\rightarrow \infty}  P_{\beta_0}\left( {\widehat{\rm LR}}(\beta_0) \geq t \middle| \widehat{Q}_R = q_R, \mathbf{v}= \mathbf{0}\right) \\
= &\ K_4 \cdot \int_{-1}^1 \left\{P_{q_U \sim \chi^2 (p)} \left( q_U \geq \frac{q_R + t}{1 + q_R u_2^2 / t} \middle| \widehat{d}_0 q_U + \widehat{d}_1 u_2 \sqrt{q_R} \sqrt{q_U} + \widehat{d}_2 q_R \leq \lambda^2 \right) \cdot (1-u^2_2)^{(p-3)/2} \right\}d u_2
\end{align*}
where 
\begin{align*}
\widehat{d}_0 &= \left(\frac{\widehat{\Omega}_{12} - \beta_0 \widehat{\Omega}_{22}}{\sqrt{b_0^T \widehat{\Omega} b_0}}\right)^2, \quad{} \widehat{d}_1 = 2 \left(\frac{\widehat{\Omega}_{12} - \beta_0 \widehat{\Omega}_{22}}{\sqrt{b_0^T \widehat{\Omega} b_0}}\right) \left(\frac{1}{\sqrt{a_0^T \widehat{\Omega}^{-1} a_0}}\right), \quad{} \widehat{d}_2 = \left(\frac{1}{\sqrt{a_0^T \widehat{\Omega}^{-1} a_0}}\right)^2 \\
K_4 &= \Gamma(p/2) /  [\pi^{1/2} \Gamma \{ (p-1)/2\} ]
\end{align*}
and $P_{q_U \sim \chi^2 (p)} (\cdot \mid E)$ is the conditional probability of $q_U$ which follows $\chi^2 (p)$.
\item[(ii)] Let $\mathcal{P} (t) = P_{\beta_0}\left( {\widehat{\rm LR}}(\beta_0) \geq t \middle| \widehat{Q}_R = q_R, \mathbf{v}= \mathbf{0}\right)$. Under the assumptions in Lemma \ref{thm:linear}, we have 
$$
 \lim_{n\rightarrow \infty} \sup_{t \in [0,1]} \left| P_{\mathbf{F}_{n}^c} \left[ \mathcal{P}\{\widehat{{\rm LR}}(\beta_0)\} \leq t \right] - t \right| = 0
$$
\end{enumerate}
\end{theorem}
The integral in Theorem \ref{cr:Q_pval} can be computed using numerical solvers (i.e. Simpson's rule), although special care is needed for $p=2$ and $p=4$ case for numerical stability; see  \cite{andrews2007performance} for details.

\subsection{Extension: Inference After Selecting Instruments via Lasso} \label{sec:extensions}
In this section, we extend our method to different conditioning events beyond the F-test. The extension is inspired by recent work of \citet{belloni2012sparse} where the authors selected instruments based on the solution to a Lasso regression between $D_i$ and $Z_i$ \citep{tibshirani_regression_1996}
$$
\widehat{\bm{\gamma}}_{\rm L} \in \amin{\bm{\gamma}} \frac{1}{2} ||\mathbf{D} - Z \bm{\gamma}||_2^2 + \lambda ||\bm{\gamma}||_1
$$
and the selected instruments is the support of $\widehat{\bm{\gamma}}_{\rm L}$. While the Lasso procedure above is not a formal test, the procedure achieves a similar goal as passing an F-test for instrument strength in that the Lasso chooses a set of relevant instruments, similar to how the F-test ``chooses'' instruments that pass the threshold $C_0$. In both cases, not taking the instrument selection process done a priori when testing $H_0$ can lead to a similar type of bias. 

Thankfully, our framework above can be adopted to handle Lasso-based pret-testing. In particular, similar to the randomized F test in Section \ref{sec:rand_F}, we can consider the randomized Lasso regression
\begin{equation} \label{eq:rlasso}
\widehat{\bm{\gamma}}_{L} \in \amin{\bm{\gamma}} \frac{1}{2} ||\mathbf{D} - Z \bm{\gamma}||_2^2 + \lambda ||\bm{\gamma}||_1 -\bm{ \omega} \cdot \bm{\gamma}
\end{equation}
where $\bm{\omega}$ is chosen from an independent randomization distribution $g$. Denote the set of selected instruments as $E$, i.e. $E = \rm supp (\widehat{\bm{\gamma}}_{L})$ and the estimated signs of $\widehat{\bm{\gamma}}_E$ as $\widehat{s}_E = \rm sign (\widehat{\gamma}_E)$. The KKT condition of the randomized Lasso stipulates that the solution $\widehat{\bm{\gamma}}_{L}$ satisfies
\begin{align*}
& - Z^T (\mathbf{D} - Z \bm{\gamma}) + \lambda \mathbf{u} - \bm{\omega} = \bm{0}, \quad{} \rm sign (\bm{\gamma}_E) = \widehat{\mathbf{s}}_E, \quad{} \bm{\gamma}_{-E} = \bm{0}, \quad{} \mathbf{u}_E = \widehat{\mathbf{s}}_E, \quad{} ||\mathbf{u}_{-E}||_{\infty} \leq 1
\end{align*}
We can use the KKT condition and elementary change-of-variables formula to transform the density from $(\mathbf{Y},\mathbf{D},Z,\bm{\omega})$ to $(\mathbf{Y},\mathbf{D},Z,\bm{\gamma}_E, \mathbf{u}_{-E})$ and arrive at the exact conditional null distribution of $\mathbf{Y}, \mathbf{D}, Z$, similar to Section \ref{sec:select_dists}:
\begin{align*} 
&\ell_{\beta_0} (\mathbf{Y}, \mathbf{D}, Z, \bm{\omega} \mid \rm supp (\bm{\gamma}_{L}) = E,  \rm sign (\bm{\gamma}_E) = \widehat{\mathbf{s}}_E) = f_{\beta_0} (\mathbf{Y}, \mathbf{D}, Z) \cdot g\left(- Z^T \mathbf{D} + Z^T Z \cdot \bm{\gamma} + \lambda \mathbf{u}\right)\cdot \mathbb{I} (\mathcal{B}) \cdot |\mathcal{J}| \\
&\mathcal{B} = \left\{\rm sign (\bm{\gamma}_E) = \widehat{\mathbf{s}}_E, \bm{\gamma}_{-E} = \bm{0}, \mathbf{u}_E = \widehat{\mathbf{s}}_E, ||\mathbf{u}_{-E}||_{\infty}\leq 1 \right\}, \quad{}|\mathcal{J}| = \rm det (Z^T Z)
\end{align*}
Here, the Jacobian $\mathcal{J}$ is from the change-of-variables formula and $\mathcal{B}$ is the conditioning event in terms of the optimization variables. If the density $f_{\beta_0}(\mathbf{Y},\mathbf{D},Z)$ is known, we can sample $\mathbf{Y},\mathbf{D},Z$ using an MCMC sampler and plug it into any test statistic to obtain the conditional null distribution of the test statistic after selecting instruments from the first-stage randomized Lasso regression and observing the estimated coefficients' signs. However, since $f_{\beta_0}(\mathbf{Y},\mathbf{D},Z)$ is unknown, we can use the asymptotics in Section \ref{sec:example_asymp_dists} and apply Lemma \ref{thm:linear}. As an example, Theorem \ref{thm:lasso} derives the asymptotic conditional null distribution of the TSLS test statistic after the instruments have been selected by the Lasso.

\begin{theorem}[Asymptotic Conditional P-Value of TSLS After Lasso]  \label{thm:lasso} 
Assume $\lambda$ is fixed in \eqref{eq:rlasso}. Let $\mathbf{S}_L = Z^T \mathbf{D}$ and $\mathbf{O}_L = \mathbf{S}_L - \widehat{\mathbf{W}}_{\mathbf{S},T} \widehat{W}_T^{-1} \cdot T$. Consider the following conditional density of the TSLS test statistic $T_{\rm TSLS}$ conditional on the support of the selected $\widehat{\bm{\gamma}}_L$, its sign, and the sufficient component
\begin{align*}
&\ell_{\beta_0} (T_{\rm TSLS}, \bm{\gamma} \mid \rm supp (\bm{\gamma}_{L}) = E,  \rm sign (\bm{\gamma}_E) = \widehat{\mathbf{s}}_E, \mathbf{O}_L) \\
&= C_{T, L} \phi_{[0,\widehat{W}_{T}]} (T_{\rm TSLS}) \cdot g\left(-\widehat{\mathbf{W}}_{\mathbf{S},T} \widehat{W}_T^{-1} \cdot T_{\rm TSLS} + Z^T Z \cdot \bm{\gamma} + \lambda \mathbf{u} - \mathbf{O}_L \right) \cdot |\mathcal{J}| \cdot \mathbb{I} (\mathcal{B}) 
\end{align*}
$$
\widehat{W}_T = 1, \quad{} \widehat{\mathbf{W}}_{\mathbf{S},T} = \frac{ \widehat{\Sigma}_{12} Z^T \mathbf{D}}{\sqrt{\widehat{\Sigma}_{11}}\sqrt{\mathbf{D}^T P_Z \mathbf{D}}}
$$
where $C_{T, L}$ is a normalizing constant. Define $\widehat{\mathcal{P}}(t) = P_{\bar{\ell}_{\beta_0}} (T_{TSLS} \geq t)$ where $\bar{\ell}_{\beta_0} = \int_{\bm{\gamma'}} \ell_{\beta_0} \{T_{\rm TSLS}, \bm{\gamma}' \mid \rm supp (\bm{\gamma}'_{L}) = E,  \rm sign (\bm{\gamma}'_E) = \widehat{\mathbf{s}}_E, \mathbf{O}_L\} $ is the marginal of $\ell_{\beta_0} \{T_{\rm TSLS}, \bm{\gamma} \mid \rm supp (\bm{\gamma}_{L}) = E,  \rm sign (\bm{\gamma}_E) = \widehat{\mathbf{s}}_E, \mathbf{O}_L\}$. Then, we have 
\begin{equation*} \label{eq:asymp_p} 
\forall \delta>0, \quad \lim_{n\rightarrow \infty} P_{\mathbf{F}_{n}^c} \left\{ \sup_{t\in \mathbb{R}} |\widehat{\mathcal{P}}(t) - \mathcal{P}(t)| > \delta\right\} = 0 \text{ and } \lim_{n\rightarrow \infty} \sup_{t \in [0,1]} \left| P_{\mathbf{F}_{n}^c} \{ \widehat{\mathcal{P}}(T) \leq t \} - t \right| = 0
\end{equation*}
\end{theorem}
As Theorem \ref{thm:lasso} illustrates, Lemma \ref{thm:linear} provides a general framework to test $H_0$ after conditioning on events that can be framed as a solution set to a convex optimization problem. So long as the assumptions in Lemma \ref{thm:linear} are reasonable, we can directly use it to construct conditional null densities of test statistics.

\section{Simulation}  \label{sec:simulation}
We conduct simulation studies to assess the proposed method in finite sample. First, we evaluate whether the null distribution of the conditional p-values is uniform; if the conditional p-values are uniformly distributed, then the conditional Type I error is controlled and consequently, the confidence interval would have the desired conditional coverage. Second, we compare our confidence interval to the {\emph{naive}} confidence interval from the TSLS test statistic that does not condition on passing the F-test. As described in Section \ref{sec:prob_statement}, our $1-\alpha$ conditional confidence interval will also have the desired $1-\alpha$ marginal coverage. In contrast, the naive interval only has the desired $1-\alpha$ coverage if no pre-testing was done. 

Formally, we generate data according to the model in \eqref{eq:model} with the following parameters: $n=1000$, $p=10$, $\beta^* = 1$, $\Sigma^*_{11} = 1$, and $ \Sigma^*_{22} = 1$. The instruments are generated according to a standard multivariate Normal, $\mathbf{Z}_i \sim N(0, I_p)$. We set the pre-test threshold to be $C_0 = 10$. Following \citet{markovic_bootstrap_2017}, our randomization term $\omega$ in \eqref{eq:randmodel} is set to be Gaussian with the standard deviation set to $0.5\sqrt{n/(n-1)} \cdot \rm std \left[ (Z^T Z)^{-\frac{1}{2}} Z^T \mathbf{D} \right]$; here $\rm std$ is an operator that computes the standard deviation of the vector inside it. at the end, the standard deviation of $\bm{\omega}$ is typically around 0.5. We vary the strength of the instrument by varying the value $r$ in $\gamma_j^* = r$ for every $j$; $r$ ranges from $0.08$ to $1$. We vary the endogeneity level by varying $\Sigma_{12}^*$ from $0$ to nearly $1$.

\begin{figure}[!ht] 
\begin{center} 
\centerline{\includegraphics[width=1.\columnwidth]{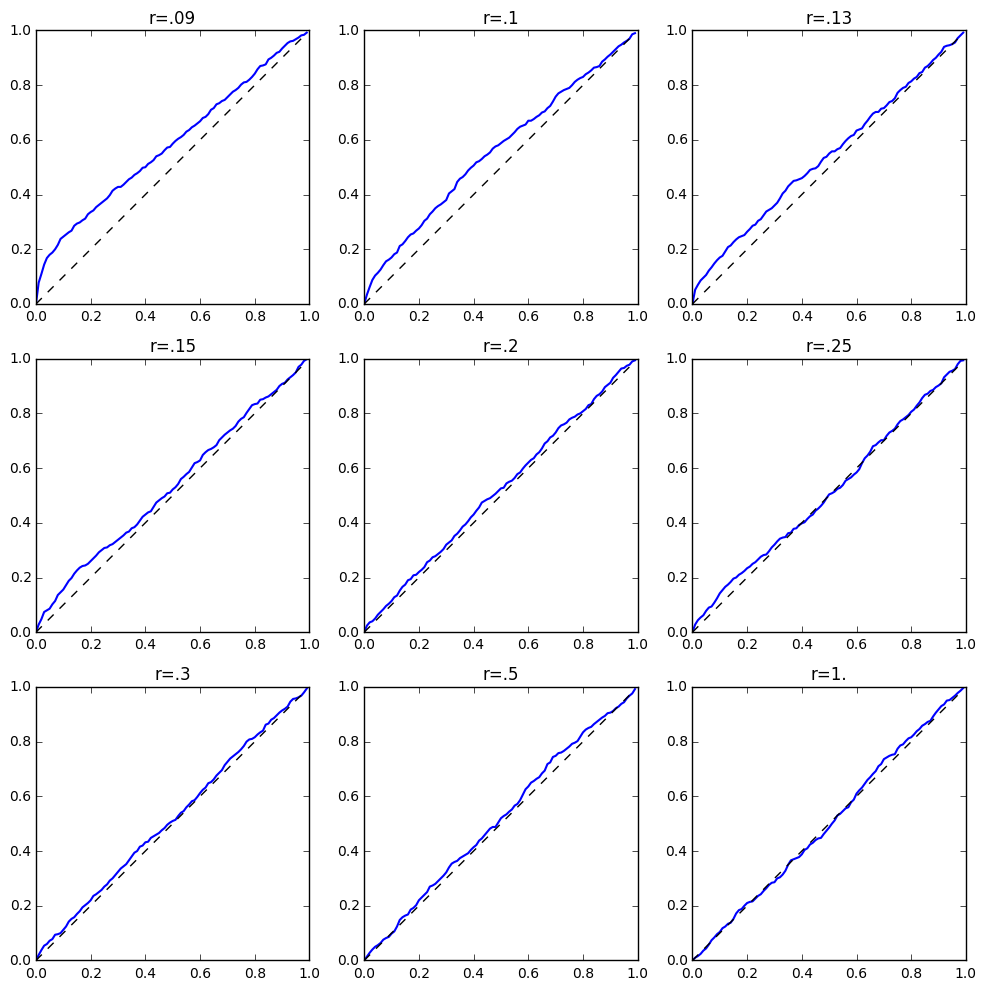}}
\caption{Comparison of CDFs between the conditional p-value of TSLS statistic and the uniform distribution. $r$ represents instrument strength as measured by setting $\gamma_j^* = r$ for all $j$. We set $\Sigma_{12}^* = 0.8$.}
\label{fig:null}
\end{center}
\end{figure}

Figure \ref{fig:null} presents the empirical cumulative density functions (CDFs)  of the TSLS' conditional p-values under $H_0: \beta^* = 1$ and $\Sigma_{12}^* = 0.8$. Each plot in Figure \ref{fig:null} varies the degree of IV strength. Except for the case when the instrument is weak at $r \leq 0.2$, the distribution of the conditional p-values is close to uniform (i.e. lines up along the 45-degree line), agreeing with what we expect from theory. When instruments are weak, the empirical distribution deviates from the uniform distribution and the conditional TSLS distribution suffers from weak instrument bias.

\begin{figure}[!ht]
\begin{center} 
\centerline{\includegraphics[width=1.\columnwidth]{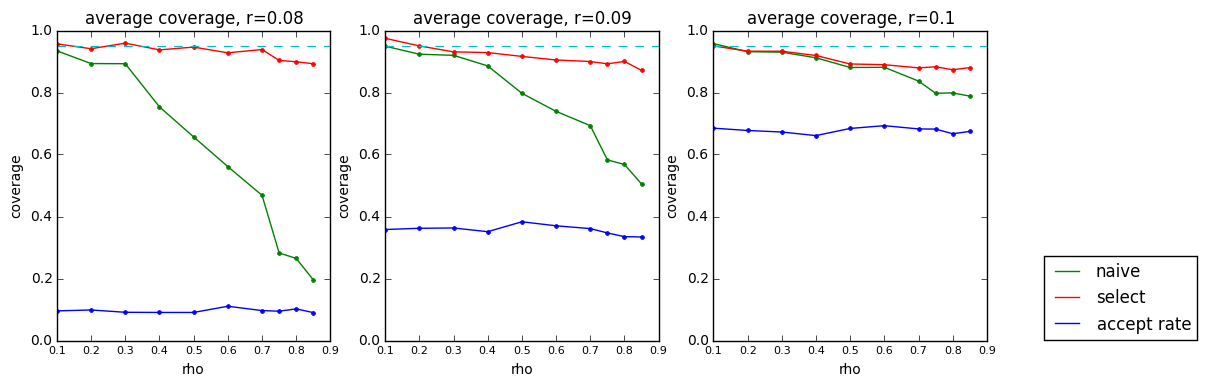}}
\caption{Comparison between the naive and the conditional confidence interval using the TSLS statistic when the instruments passed the F-test. The x-axis represents endogeneity measured by $\Sigma_{12}^*$. Each plot from left to right increases instrument strength by setting $r =0.08, 0.09$, and $r=0.10$. The green and red lines show the coverage rates of naive and conditional confidence intervals. The blue line shows the proportion of times the instruments passed the F-test pre-test.}
\label{fig:ci2}
\end{center}
\end{figure}

Figure \ref{fig:ci2} presents the coverage rate of the conditional confidence interval versus the naive confidence interval based on the TSLS test statistic when the instruments pass the F-test pre-test. We also plot the ``passing'' rate of the F-test pre-test, i.e. how many times did the instruments have F-statistics above $C_0 = 10$. The setting varies instrument strength from $r = 0.08$ to $r=0.1$. While both intervals start to suffer from coverage as endogeneity parameters $\Sigma_{12}^*$ increases, especially around $0.9$, we see that our conditional intervals always maintain close to nominal 95\% marginal coverage compared to the naive interval, whose coverage can drop to 20\% in extreme cases. We also notice that as the passing rate of the pre-test increases, the gap between the naive and conditional confidence interval narrows. This suggests that the effect of not accounting for the F-test is most severe if the passing rate is low, or roughly if the instruments are not strong. 

\begin{figure}[!ht]
\begin{center} 
\centerline{\includegraphics[width=1.\columnwidth]{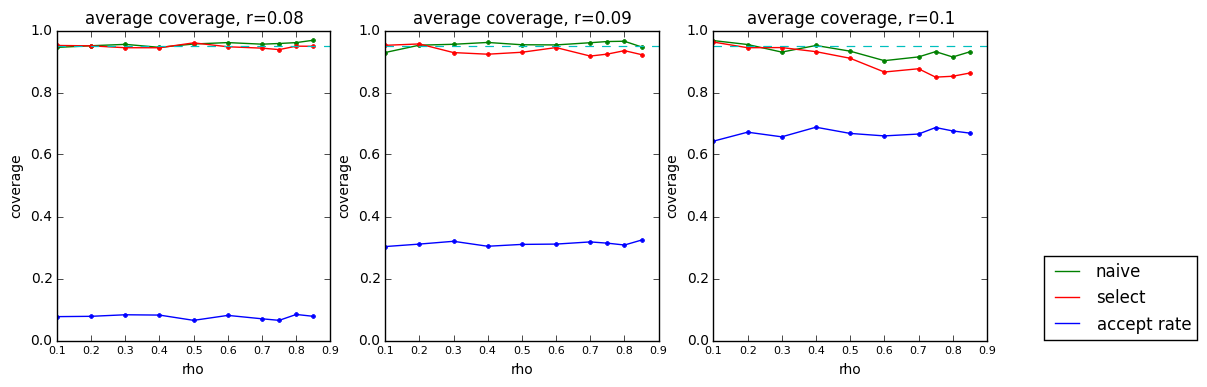}}
\caption{Comparison between the naive and conditional confidence interval using the CLR test when the instruments did not pass the F-test. The x-axis represents the endogeneity level measured by $\Sigma_{12}^*$. Each plot from left to right increases instrument strength by setting $r =0.08, 0.09$, and $r=0.10$. The green and red lines show the coverage rates of naive and conditional inference. The blue line shows the proportion of times the instruments passed the F-test pre-test.}
\label{fig:ci4}
\end{center}
\end{figure}

Figure \ref{fig:ci4} presents the coverage rate of the conditional confidence interval versus the naive confidence interval based on the CLR test statistic when instruments failed to pass the pre-test. Like Figure \ref{fig:ci2}, we also plot the passing rate of the pre-test with the threshold $C_0 = 10$ and vary the instrument strength from $r = 0.08$ to $r=0.1$. Overall, we find that the confidence interval based on the CLR test conditional on not passing the pre-test is nearly identical to the usual confidence interval without conditioning on the pre-test; both intervals are very wide and sometimes infinite. This suggests that for the CLR test and in regimes where the instruments are weak, the difference between accounting for the pre-test and not accounting for it is minimal and the practical difference between using unconditional or conditional tests is negligible. However, as the instruments' strength increase and the F-test gets closer to the $C_0$ boundary, the effect of not conditioning on the pre-test begins to appear, with the naive CI's coverage starting to dip below nominal levels whereas the conditional CI's coverage remains closer to 95\%.

In the supplementary materials, we conduct additional simulation studies for different test statistics. In particular, we also verify whether another weak-instrument robust test, the Anderson-Rubin test, shows a similar pattern as the CLR test. In short, we find that for the Anderson-Rubin test, the effect of not conditioning on pre-testing is minimal when instruments are weak.

\section{Applications} 
We demonstrate our proposed method on two datasets. The first dataset concerns instruments that passed the F-test pre-test at the usual threshold value of $C_0 = 10$ and the second dataset concerns instruments that do not pass the F-test pre-test, leading us to use the weak-instrument robust CLR test statistic. Both datasets study the same effect, the effect of years of schooling on earnings, but the first dataset uses instruments proposed by \citet{card1993using} while the second dataset uses instruments proposed by \citet{angrist1991does}. As we will see in both examples, by taking into account the pre-test, we may arrive at different conclusions about the treatment effect compared to not taking the pre-test into account. More broadly, the two empirical examples highlight the importance of taking instrument-strength tests into consideration for a more honest statistical assessment of the treatment effect.

\subsection{Geographical Variation as Instruments} \label{sec:app_geo}
The first dataset is from \citet{card1993using} where the author explored the use of geographic variation, specifically college proximity, as an instrument to estimate the causal effect of years of schooling on earnings. His analysis uses $n=3010$ individuals from the National Longitudinal Survey of Young Men (NLSYM). NLSYM began in 1966 with 5525 men around the age of 14 and 24 and the survey continued to track these men through 1981; the dataset can be found on the author's website. \citet{card1993using}'s motivation for using geographical proximity as an instrument was based on an observation that men who grew up in labor markets with a college nearby had significantly higher education levels than men who did not grow up near a college. For our application, we replicate \citet{card1993using}'s model 1 and model 2 in Table 3 of panel A. We re-run Card's analysis to verify our replication process. We also re-run Card's analysis after accounting for testing for instrument strength.

Both of Card's models have completed years of education at 1976 as the treatment, log hourly wages in 1976 as the outcome, and a binary instrument that indicates whether the subject lived near a college in 1966. Both models control for the following exogenous variables: race (black or not), an indicator for southern residence (south or not), subject's residence in a standard metropolitan statistical area (SMSA) (yes or no), 8 indicators for subject's regional background, and experience in years and as well as its squared. Model 2 controlled for 14 family-specific covariates, including mother's and father's education, indicators for missing father's or mother's education, interactions of mother's and father's education, and dummies representing family structure at age 14. Model 1 did not control for these 14 family-specific covariates. See \citet{card1993using} for detailed explanations of each variable.

Table \ref{tab:card} summarizes the result. The first three columns represent Card's original analysis under the two models. As \citet{card1993using} concluded, the effect of years of schooling on earnings is significant using TSLS with college proximity as an instrument. For TSLS, they found the p-value of the null hypothesis of no effect to be around $0.01$ under both models, with 95\% confidence intervals ranging from $0.024$ to $0.25$. \citet{card1993using} also assessed instrument strength prior to producing these inferential results by using the F-test and found that the F-statistic was around 14, which is above the usual acceptable threshold of $C_0 = 10$.

\begin{table}[!ht]
\caption{Result from \citet{card1993using} using college proximity as an instrument. The first three columns replicated Card's analysis in Table 3, Panel A and each row represents different model specifications in Panel A. Columns ``Conditional 95\% CI'' and ``Conditional P-value'' represent the analysis of the treatment effect after conditioning on passing the F-test at $C_0 = 10$. The TSLS test statistic are used to test for the treatment effect. The last column represents the F-statistic for instrument strength. Standard errors are in parenthesis.} \label{tab:card}
\begin{center}
\begin{tabular}{|c c p{2cm} c c c c|}
\hline
Model & TSLS & 95\% CI & P-value & Conditional 95\% CI  & Conditional P-value & F-test \\
\hline
Model 1 & 0.132 (0.055) & [0.024, 0.239] & 0.016 & [-0.094, 0.357] & 0.602 & 13.322 \\
Model 2 & 0.140 (0.055) & [0.033, 0.248] & 0.011 & [-0.072, 0.327] & 0.811 & 15.081 \\
\hline
\end{tabular}
\end{center}
\end{table}

The next set of columns in Table \ref{tab:card} computes conditional confidence intervals and p-values where we test the effect of education after seeing that the instrument had an F-statistic above $C_0 = 10$. We see that after conditioning on the pre-test, the treatment effect is no longer significant at $\alpha = 0.05$ and the confidence interval for the education effect is larger. The conditional p-values range from 0.6 to 0.8, nowhere as significant as the analysis that did not condition on the pre-test done a priori. Also, the 95\% conditional confidence intervals range from $-0.1$ to $0.36$, with more mass towards the positive values of the interval.  In other words, ignoring that the data was used twice, one for testing instrument's strength to select a strong IV and the other to test the treatment effect, led to more optimistic conclusions about the effect of schooling on earnings.

\subsection{Quarter of Birth as Instruments}
The second dataset is from \citet{angrist1991does} where the authors proposed to use quarter of subject's birth as an instrument to estimate nearly identical effects as above, the effect of schooling on earnings. The authors' analysis used $n = 1,063,634$ men from the U.S. Census; the dataset can be found on the first author's website. The authors argued that the quarter of birth is related to educational attainment because certain school policies allowed individuals who were born earlier in the school year to start school earlier than those born near the end of the school year. Using the birth quarter instrument, they found a significant effect of education on earnings that was similar in magnitude to the analysis by \citet{card1993using}.

One of the criticisms of \citet{angrist1991does}'s analysis is that the birth quarter instrument was too weak to draw meaningful inference, especially using the TSLS estimator \citep{bound1995problems}. Indeed, in our replication analysis below, we see that all the F-tests for instrument strength under different model specifications are well below the usual threshold $C_0 = 10$. Based on the pre-test results, we use the weak-instrument robust CLR test statistic, specifically computing the conditional distribution of the CLR statistic when the F-statistic failed to exceed the threshold $C_0 = 10$. Our findings are in Table \ref{tab:angrist}. We focus on model (8) across Tables IV, V, and VI in \citet{angrist1991does} since this controls for the full set of exogeneous variables and put the result of models (2), (4), and (6) in the supplementary materials.

\begin{table}[!ht]
\centering
\caption {Result from \citet{angrist1991does} using quarter of birth as instruments. Column ``CLR'' replicates the authors' analysis, but with the CLR test statistic. Columns ``Conditional 95\% CI'' and ``Conditional P-value'' represent the analysis of the treatment effect after conditioning on not passing the F-test at $C_0 = 10$. he last column represents the F-statistic for instrument strength.}
\label{tab:angrist}
%\begin{tabular}{|c|c|c|c|c|p{2.5cm}|p{2cm}|c|}
\begin{tabular}{| c c c c c c |} \hline
Model & CLR 95\% CI & CLR P-value & Conditional 95\% CI & Conditional P-value & F-test \\ \hline
(IV)(8) & [-$\infty$, +$\infty$] & 0.3717 &  [-$\infty$, +$\infty$] & 0.3717 & 0.9595 \\
(V)(8) & [-0.8312, 0.7532] & 0.4254 &  [-0.8312, 0.7532] & 0.4254 & 1.5071 \\
(VI)(8) & [0.0238, 0.2671] & 0.0182 & [0.0238, 0.2671] & 0.0182 & 2.5557 \\ \hline
\end{tabular}
%Model & TSLS & 95\% CI & P-value & CLR 95\% CI & CLR P-value & Conditional 95\% CI & Conditional P-value & F-test \\ \hline
%(IV)(8) & 0.1007 (0.0334) & [0.0352, 0.1660] & 0.0026 & [-$\infty$, +$\infty$] & 0.3717 &  [-$\infty$, +$\infty$] & 0.3717 & 0.9595 \\
%(V)(8) & 0.0600 (0.0299) & [0.0031, 0.1167] & 0.0387 & [-0.8312, 0.7532] & 0.4254 &  [-0.8312, 0.7532] & 0.4254 & 1.5071 \\
%(VI)(8) & 0.0779 (0.0239) & [0.0308, 0.1243] & 0.0012 & [0.0238, 0.2671] & 0.0182 & [0.0238, 0.2671] & 0.0182 & 2.5557 \\ \hline
\end{table}

The first three columns replicate the original analysis done by \citet{angrist1991does} in Tables IV, V, and VI, but with the CLR test. Specifically, Table IV computed the CLR confidence intervals of education's effects among men born between 1920 and 1929 ($n = 247,199$). %1970 census
Treatment was years of education. Instruments were a full set of quarter-of-birth times year-of-birth interactions. Outcome was log weekly earnings. The models included the following exogenous variables: race indicator (black or not), SMSA indicator, marriage status indicator, 8 regional dummies, 9 year-of-birth dummies, age, and age squared. Table V used the same instrument, treatment, outcome, and exogenous covariates, but focusd on men born between 1930 and 1939 ($n = 329,509$). Table VI used the same variables except that they are men born between 1940 and 1949 ($n = 489,926$).

The columns labeled ``conditional'' in Table \ref{tab:angrist} presents our conditional method; we use $[-\infty, +\infty]$ to denote possibly infinite 95\% conditional confidence intervals if, from our sampling procedure above, we obtain intervals $[-M, +M]$ with $M \geq 10^5$. The next two columns presents the CLR p-value and confidence interval without conditioning. The last column represents the F-statistic used in the pre-test. We see that the unconditional CLR p-values and confidence intervals that do not take into account the F-test pre-test are numerically similar to the conditional ones across the different models. Overall, the data analysis reinforces the observations we noticed in our simulation study concerning the CLR test, specifically that weak-instrument robust tests with very weak instruments are insensitive to conditioning on the F-test.

\section{Conclusion}  \label{sec:conclusion}
In this paper, we propose to test for the treatment effect after testing for instrument strength with an F-statistic. We propose a sampling based method that controls the Type I error conditional on passing (or not passing) the pre-test and produces confidence intervals based on these conditioning events. We also extend our method to handle weak instruments and other types of conditioning events, most notably based on the Lasso. We show through simulation studies that our conditional CIs achieves nominal coverage levels whereas the naive CIs without correcting for the pre-test step often has lower-than-expected coverage. Finally, we re-examine two studies in labor economics that estimate the return on education. We show that a naive analysis of the treatment effect without taking into account the pre-test over-inflates the statistical significance of the treatment effect. We believe the empirical example, especially the re-analysis of Card's result, highlights one severe consequence of not accounting for instrument strength testing and should serve as a cautionary reminder for investigators working with instruments. More broadly, we believe that our application of selective inference is a viable path to resolve inferential issues conditional on running diagnostic or specification/robustness checks; see \citet{bi2019inference} for another example of selective inference in IV settings.

%\section*{Acknowledgements}\label{acknowledgements}

%Acknowledgements should include contributions from anyone who does not
%meet the criteria for authorship (for example, to recognize
%contributions from people who provided technical help, collation of
%data, writing assistance, acquisition of funding, or a department
%chairperson who provided general support), as well as any funding or
%other support information.

%\section*{Conflict of interest}

\selectlanguage{english}
\FloatBarrier
\bibliography{selectIV.bib}

\end{document}